\documentclass[aps,floatfix,showpacs,preprintnumbers,amsmath,amssymb]{revtex4}
\usepackage{natbib}
\usepackage{dcolumn}
\usepackage{graphicx}
\usepackage{enumerate}
\usepackage{hyperref}
\newcommand{\mnras}{Mon.~Not.~R.~Astron.~Soc.}

\newcommand{\phrd}{Phys.~Rev.~D.}
\newcommand{\jcap}{J.~Cosmol.~Astropart.~Phys.}
\newcommand{\be}{\begin{equation}}
\newcommand{\ee}{\end{equation}}
\newcommand{\bea}{\begin{eqnarray}}
\newcommand{\eea}{\end{eqnarray}}
\def\[{\begin{equation}}
\def\]{\end{equation}}
\begin{document}
\title{Testing General Relativity at Cosmological Scales: Implementation and Parameter Correlations}
\author{Jason N.  Dossett$^1$\footnote{Electronic address: jdossett@utdallas.edu}, Mustapha Ishak$^1$\footnote{Electronic address: mishak@utdallas.edu}, Jacob Moldenhauer$^2$\footnote{Electronic address: JMoldenhauer@fmarion.edu}}
\affiliation{
$^1$Department of Physics, The University of Texas at Dallas, Richardson, Texas 75083, USA;
$^2$Department of Physics and Astronomy, Francis Marion University, Florence, SC 29506, USA;}
\date{\today}
%
%\tableofcontents
%
\begin{abstract}
The testing of general relativity at cosmological scales has become a possible and timely endeavor that is not only motivated by the pressing question of cosmic acceleration but also by the proposals of some extensions to general relativity that would manifest themselves at large scales of distance. 
We analyze here correlations between modified gravity growth parameters 
and some core cosmological parameters using the latest cosmological data sets including the refined Cosmic Evolution Survey 3D weak lensing.   
We provide the parametrized modified growth equations and their evolution. We implement known functional and binning approaches, and propose a new hybrid approach to evolve the modified gravity parameters in redshift (time) and scale. 
The hybrid parametrization combines a binned redshift dependence and a smooth evolution in scale avoiding a jump in the matter power spectrum. 
The formalism developed to test the consistency of current and future data with general relativity is implemented in a package that we make publicly available and call \texttt{ISiTGR} ({\it Integrated Software in Testing General Relativity}), an integrated set of modified modules for the publicly available packages \texttt{CosmoMC} and \texttt{CAMB}, including a modified version of the integrated Sachs-Wolfe-galaxy cross correlation module of Ho et al and a new weak-lensing likelihood module for the refined Hubble Space Telescope Cosmic Evolution Survey weak gravitational lensing tomography data.  
We obtain parameter constraints and correlation coefficients finding that modified gravity parameters are significantly correlated with $\sigma_8$ and mildly correlated with $\Omega_m$, for all evolution methods. The degeneracies between $\sigma_8$ and modified gravity parameters are found to be substantial for the functional form and also for some specific bins in the hybrid and binned methods indicating that these degeneracies will need to be taken into consideration when using future high precision data. 
\end{abstract}
\maketitle
\section{introduction}
There has been a growing interest in testing general relativity (GR) at cosmological scales. This has been triggered by the quest to understand the origin of cosmic acceleration which is one of the biggest puzzles in cosmology today.  Specifically, one would like to test whether cosmic acceleration is due to some unknown, "dark energy" component in the Universe or rather an extension or modification to gravity physics on cosmological scales. The growth rate of large scale structure proved to be a solid discriminator between these two possible explanations for cosmic acceleration and has been studied within this context by a number of papers.  For example, some early papers explored inconsistencies between parameter constraints derived using observations that depended primarily on the growth rate of structure versus constraints derived from observations that probed only the expansion history. The presence of such inconsistencies would call into question the underlying gravity theory (see \cite{ScoccimarroStarkman2004,Song2005a,Ishak2005,KnoxSongTyson2006}). These were followed by a number of papers using certain growth parameters that take distinct values for different gravity theories (see \cite{Linder2005, ZhangEtal2007, HuandSawicki2007, CaldwellCoorayandMelchiorri2007, AcquavivaEtAl2008,KunzandSapone2008, Gong2008, Daniel2008, GongIshakWang2009, BertschingerandZukin2008, HutererandLinder2007, KoivistoandMota2006, GannoujiMoraesandPolarski2009,Koyama, Zhang2006, LinderCahn, Song2007, Dore, Gabadadze, Polarski, Hu2008, Jain, Wei, Dent2009, Fu, Ishak2009, Linder2009, Serra, Song2009b, Song2009c, Thomas, Tsujikawa, Wu, GongBo2009a, GongBo2009b, Acquaviva2010, Dossett, Ferreira, Jing, Pogosian2010, Song2010}). In this second approach, some parameters characterizing the growth rate are introduced in order to parametrize deviations from GR. These parameters have known values in GR but different values in modified gravity theories. So one possible goal is to constrain these parameters, in addition to the usual cosmological parameters, and see if they are consistent or not with the GR values.  Also, recently Basilakos {\it et al.} proposed using the evolution of the linear bias as a way to test for deviations from GR \cite{Basilakos}.

These approaches have been applied using current available data sets and future simulated ones and in order to constrain growth parameters related to modifications of the perturbed Einstein equations in an effort to look for any indications of deviations from GR, but so far, none have been detected \cite{WangEtal2007, IshakReport, Daniel2009, Bean2010, Daniel2010, DL2010, GongBo2010, Lombriser, Toreno,Dossett2011,MGCAMB2}.  

In this paper, we study correlations between modified gravity parameters and 
core cosmological parameters. We implement the evolution in time and scale of the modified gravity parameters using a functional form method and a binning method. We also propose and implement a new hybrid method that combines both. In the hybrid method, the evolution in time is represented in two redshift bins while the scale evolution follows a monotonic functional form. 
This provides a smooth evolution in scale combined with a binned redshift (time) dependence that was shown to be more robust than time functional forms as we discuss in the next section, see also \cite{Dossett2011,GongBo2010,SongZhao2011}. We also describe the numerical framework  that we introduce here as the package: {\it \textbf{I}ntegrated \textbf{S}oftware \textbf{i}n \textbf{T}esting \textbf{G}eneral \textbf{R}elativity}, \texttt{ISiTGR} (pronounced {\it Is it GR} and available publicly at \url{http://www.utdallas.edu/~jdossett/isitgr/}).  \texttt{ISiTGR} is an integrated set of modified modules for the publicly available packages \texttt{CosmoMC} \cite{cosmomc} and \texttt{CAMB} \cite{LewisCAMB}.  It combines all the modifications to those packages and a modified version to the of the Integrated Sachs-Wolfe (ISW)-galaxy cross correlations module by Ho {\it et al.} \cite{ISWHo,ISWHirata} to test GR. We also include our weak-lensing likelihood module for the recently refined Hubble Space Telescope (HST) Cosmic Evolution Survey (COSMOS) weak-lensing tomography analysis in \cite{Schrabback2010} which has also been modified to test GR, and a new baryon acoustic oscillation (BAO) likelihood module for the recently released WiggleZ Dark Energy Survey BAO measurement data \cite{BlakeBAO}.
%
%
%%%%%%%%%%%%%%%%%%%%%%%%%%%%%%%%%%%%%%%%%%%%%%%%%%%%%%%%%%%%%%%%%%%%%%%%%%%%%%%%%%%
\section{Parametrizing Deviations of the Growth Equations from General Relativity}
%%%%%%%%%%%%%%%%%%%%%%%%%%%%%%%%%%%%%%%%%%%%%%%%%%%%%%%%%%%%%%%%%%%%%%%%%%%%%%%%%%%
%
%%%%%%%%%%%%%%%%%%%%%%%%%%%%%%%%%%%%%%%%%%%%%%%%%
\subsection{Growth Equations in General Relativity}
%%%%%%%%%%%%%%%%%%%%%%%%%%%%%%%%%%%%%%%%%%%%%%%%%
%%%%%%
In the conformal Newtonian gauge the perturbed Friedmann-Lemaitre-Robertson-Walker (FLRW) metric is written as
\be
ds^2=a(\tau)^2[-(1+2\psi)d\tau^2+(1-2\phi)dx^idx_i],
\label{eq:FLRW}
\ee
where $a(\tau)$ is the scale factor normalized to 1 today, the $x_i$'s are the comoving coordinates, and $\tau$ is conformal time.  $\phi$ and $\psi$ are scalar potentials describing the scalar mode of the metric perturbations.  

Using the first-order perturbed Einstein equations, while working in Fourier $k$ space, we can get two very useful equations that describe the evolution of the scalar potentials.  The combination of the time-time and time-space equations gives the Poisson equation describing the potential $\phi$.  Then, to relate the two potentials to one another we take the traceless, space-space component of these equations.  Explicitly, these equations are
\bea
k^2\phi  &=&-4\pi G a^2\sum_i \rho_i \Delta_i
\label{eq:P}\\
k^2(\psi-\phi) &=& -12 \pi G a^2\sum_i \rho_i(1+w_i)\sigma_i,
\label{eq:2E}
\eea
where $\rho_i$ and $\sigma_i$ are the density and the shear stress, respectively, for matter species, $i$.  
$\Delta_i$ is the gauge-invariant, rest-frame overdensity for matter species, $i$, the evolution of which describes the growth of inhomogeneities.  It is defined by
\be
\Delta_i = \delta_i +3\mathcal{H}\frac{q_i}{k},
\label{eq:Deltadef}
\ee 
where $\mathcal{H} =\dot{a}/a$ is the Hubble factor in conformal time, and for species $i$, $\delta_i=\delta \rho_i/\bar{\rho}$ is the fractional overdensity and $q_i$ is the heat flux and is related to the divergence of the peculiar velocity, $\theta_i$, by $\theta_i=\frac{k\ q_i}{1+w_i}$.  Enforcing the conservation of energy momentum on the perturbed matter fluid, these quantities for uncoupled fluid species or the mass-averaged quantities for all the fluids evolve as described in \cite{Ma}:
\bea
\dot{\delta} & = & -k q +3(1+w)\dot{\phi}+3\mathcal{H}(w-\frac{\delta P}{\delta\rho})\delta
\label{eq:deltaevo}\\
\frac{\dot{q}}{k}&= &-\mathcal{H}(1-3w)\frac{q}{k}+\frac{\delta P}{\delta\rho}\delta+(1+w)\left(\psi-\sigma\right).
\label{eq:qevo}
\eea
Above, $w=p/\rho$ is the equation of state.  Combining these two equations, we can express the evolution of $\Delta$ by
\be
\dot{\Delta} = 3(1+w)\left(\dot{\phi}+\mathcal{H}\psi\right)+3\mathcal{H}w\Delta -\left[k^2+3\left(\mathcal{H}^2-\dot{\mathcal{H}}\right)\right]\frac{q}{k}-3\mathcal{H}(1+w)\sigma.
\label{eq:Deltaevo}
\ee

Equations (\ref{eq:P}),(\ref{eq:2E}),(\ref{eq:deltaevo}), and (\ref{eq:qevo}) are coupled to one another; combining them, along with the evolution equations for $a(\tau)$, we can describe the growth history of structures in the Universe.
%
%%%%%%%
%%%%%%%%%%%%%%%%%%%%%%%%%%%%%%%%%%%%%%%%%%%%%%%%%%%%%%%%%%%%%%%%%%%%%%%%%%%%
\subsection{Modified Gravity Growth Parameters}
%%%%%%%%%%%%%%%%%%%%%%%%%%%%%%%%%%%%%%%%%%%%%%%%%%%%%%%%%%%%%%%%%%%%%%%%
%%%%%%%%%%%
%%
Parametrizing both modifications to Poisson's equation, (\ref{eq:P}),  as well as the ratio between the two metric potentials $\phi$ and $\psi$ in the perturbed FLRW metric (called \emph{gravitational slip} by Caldwell {\it et al.} \cite{CaldwellCoorayandMelchiorri2007}) has recently been the subject of a lot of the work on testing general relativity; see, for example, \cite{CaldwellCoorayandMelchiorri2007,Daniel2009,GongBo2010,Bean2010,Daniel2010,DL2010}.  The parameters we use in this paper to describe modifications to the growth [{\it modified gravity MG parameters}] are based upon those used in \cite{Bean2010}.  

The parametrized modifications to the growth equations proposed by \cite{Bean2010} directly modify Eqs. (\ref{eq:P}) and (\ref{eq:2E}) and make no assumptions as to the time when a deviation from GR is allowed. These modifications are as follows:
\bea
k^2\phi  &=& -4\pi G a^2\sum_i \rho_i \Delta_i \,  Q
\label{eq:ModP}\\
k^2(\psi-R\,\phi) &=& -12 \pi G  a^2\sum_i \rho_i(1+w_i)\sigma_i \, Q,
\label{eq:Mod2E}
\eea
where $Q$ and $R$ are the MG parameters.  The parameter $Q$ represents a modification to the Poisson equation, while the parameter $R$ quantifies the gravitational slip (at late times, when anisotropic stress is negligible, $R=\psi/\phi$). In our code, rather than using the parameter $R$ which is degenerate with $Q$, we instead use the parameter $\mathcal{D} = Q(1+R)/2$ as suggested in \cite{Bean2010} {(this parameter is equivalent to the parameter $\Sigma$ in \cite{Song2010,GongBo2010} or $\mathcal{G}$ of \cite{DL2010}).}  Combining Eqs. (\ref{eq:ModP}) and (\ref{eq:Mod2E}), we arrive at the second modified growth equation used in this paper:
\be
k^2(\psi+\phi) = -8\pi G a^2\sum_i \rho_i \Delta_i \, \mathcal{D} \, -12 \pi G  a^2\sum_i \rho_i (1+w_i)\sigma_i \, Q.
\label{eq:ModPSum}
\ee
So, the modified growth equations are (\ref{eq:ModP}) and (\ref{eq:ModPSum}), and $Q$ and $\mathcal{D}$ are now the MG parameters.  As discussed in our previous work \cite{Dossett2011}, this approach of using the parameter $\mathcal{D}$ instead of $R$ is also useful because observations of the weak-lensing and ISW are sensitive to the sum of the metric potentials $\phi +\psi$ and its time derivative respectively. Thus observations are able to give us direct measurements of this parameter.
%%
%%%%%%%%%%%
%%%%%%%%%%%%%%%%%%%%%%%%%%%%%%%%%%%%%%%%%%%%%%%%%%%%%%%%%%%%%%%%%%%%%%%%
\subsection{Different Approaches to Evolving Modified Growth Parameters in Time and Scale}
%%%%%%%%%%%%%%%%%%%%%%%%%%%%%%%%%%%%%%%%%%%%%%%%%%%%%%%%%%%%%%%%%%%%%%%%
%%%%%%%%%%%
%%
To date, there have primarily been two approaches to evolving the MG parameters in time and scale; one using a continuous functional form and the other based on binning. {We implement in \texttt{ISiTGR} the two approaches and, additionally, a new hybrid approach, as we explain below.} 

The first approach involves defining a functional form for each parameter that allows it to evolve monotonically in both time and scale.   This allows one to make no assumptions as to when a deviation from general relativity is allowed.  Such an approach was taken in for example \cite{Bean2010}.  In that work the functional form,
\be
X(k,a) = \left[X_0 e^{-k/k_c}+X_\infty(1-e^{-k/k_c})-1\right]a^s +1,
\label{eq:BeanEvo}
\ee
was assumed,  where $X$ denotes either $Q$ or $R$ in Eqs. (\ref{eq:ModP}) and (\ref{eq:Mod2E}).  Thus a total of six model parameters are used to test GR:
$Q_0$, $R_0$, $Q_\infty$, $R_\infty$, $k_c$, and $s$.  The parameters $s$ and $k_c$ parametrize time and scale dependence respectively, with GR values $s=0$ and $k_c=\infty$.  $Q_0$ and $R_0$ are the present-day superhorizon values while $Q_\infty$ and $R_\infty$ are the present-day subhorizon values of the $Q(k,a)$ and $R(k,a)$ , all taking GR values of $1$. 

In the second approach, instead of evolving each of the parameters assuming some functional form, one can bin the MG parameters. This approach allows the parameters to take on different values in predefined redshift and scale bins.  This technique was used in, for example, \cite{GongBo2010,DL2010}.  In those works two redshift and two scale bins were defined and for redshifts above a certain critical redshift, GR was assumed to be valid.  In each bin the parameters were allowed to take on different values resulting in a total of eight model parameters used to test GR.

The third approach that we propose here is a hybrid one where 
the evolution in redshift (or time) is binned into two redshift bins, 
but the evolution in scale evolves monotonically in the same way as the 
functional form above. Our motivation for this hybrid binning approach is 
that it takes advantage of an evolution in scale that is not so abrupt as 
that in the traditional binning method, while still taking 
advantage of a redshift (time) dependence expressed in the form of bins, which 
was shown to be more robust than time functional forms \cite{Dossett2011,GongBo2010,SongZhao2011}. 
For example, in \cite{Dossett2011}, we found that binning methods do not display the extent of tensions between the MG parameters (preferred by different data sets) as in the functional form method, where tensions are exacerbated by the chosen functional form. Also, in Ref. \cite{SongZhao2011}, the author found similar to what we noticed and that the constraints on MG parameters depend strongly on the parameter $s$ (the scale factor exponent in the functional parametrization). They further looked at ways to remove this strong dependence on the parameter $s$ suggesting and exploring binning as a solution.

To take advantage of all these techniques, we have developed two versions of our code.  One version of the code uses the functional form (\ref{eq:BeanEvo}).  It provides the option to apply the functional form (\ref{eq:BeanEvo}) to either $Q$ and $R$ (as done in \cite{Bean2010}) or $Q$ and $\mathcal{D}$.  The other version of the code is based on binning methods.  It provides the option of the second approach with traditional binning (described above), or alternatively the third, hybrid approach with two redshift bins, but the evolution in scale evolves monotonically.

\begin{figure}
\begin{center}
\begin{tabular}{|c|}
\hline
{\includegraphics[width=4.in,height=2.3in,angle=0]{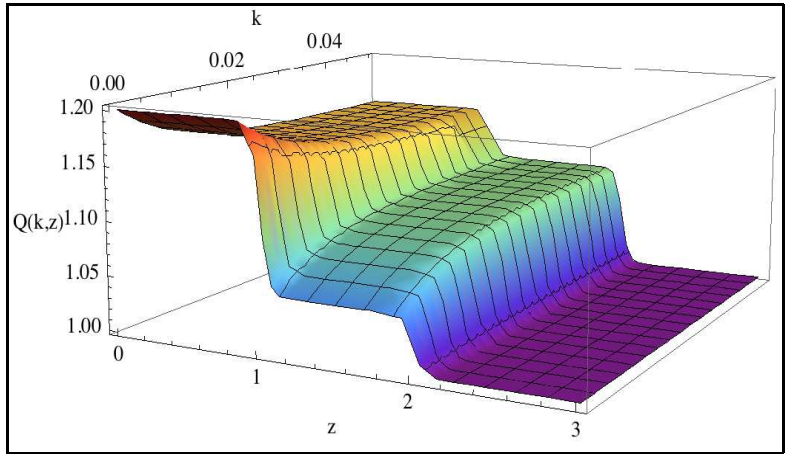}}\\
\hline  
\end{tabular}
\caption{\label{fig:binplot}
MG parameter evolution in redshift and scale modeled using a new hybrid method. We plot here a 3D representation for an example of the new hybrid binned evolution for the MG parameter $Q(k,a)$ as given by our Eqs. (\ref{eq:EvoBinZ}) and (\ref{eq:EvoBinKExp}) for the parameters $Q(k,a)$ with $Q_1\,=\, 1.20, \, Q_2\, =\, 1.15, \, Q_3\, =\, 1.05, \,Q_4\, = \, 1.10, \, z_{TGR}\, =\,2, \, \mbox{and } k_c\, =\, 0.01$.  We can see along the $z$-axis how the binned aspect can allow for different best fit values for the MG parameters in the redshift space while along the $k$-axis we can see the monotonic evolution in $k$ evolving from some large scale (small $k$) value to a small scale (large $k$) value exponentially. The hybrid parametrization combines the $z$-binning method that was shown to be robust with a smooth evolution in $k$ space.} 
\end{center}
\end{figure}

In the binning version of the code, we evolve only $Q$ and $\mathcal{D}$.  Transitions between the redshift bin are evolved following \cite{Pogosian2010,GongBo2010} and use a hyperbolic tangent function with a transition width $z_{tw} = 0.05$.  In this way the binning can actually be written functionally as (with $X$ representing $Q$ or $\mathcal{D}$)
\be 
X(k,a) =\frac{1 + X_{z_1}(k)}{2}+\frac{X_{z_2}(k) - X_{z_1}(k)}{2}\tanh{\frac{z-z_{div}}{z_{tw}}}+\frac{1 - X_{z_2}(k)}{2}\tanh{\frac{z-z_{TGR}}{z_{tw}}},
\label{eq:EvoBinZ}
\ee
where $z_{div}$ is the redshift where the transition between the two redshift bins occurs and $z_{TGR}$ is the redshift below which GR is to be tested.  We hard code $z_{TGR} = 2z_{div}$ to give us equally sized bins, but this of course is optional and can easily be changed.  $X_{z_i}(k)$ represents the binning method for $k$ in the $i$th $z$ bin.  For the suggested hybrid method it has the form 
\bea
X_{z_1}(k) &=& X_1 e^{-k/k_c}+X_2(1-e^{-k/k_c}) \label{eq:EvoBinKExp} \\ \nonumber
X_{z_2}(k) &=& X_3 e^{-k/k_c}+X_4(1-e^{-k/k_c}),
\eea
while with traditional binning in principle evolves as
\bea
X_{z_1}(k)&=&\left\{\begin{array}{ll}
X_1 & \mbox{if }k<k_c\\
X_2 & \mbox{if }k\geq k_c,
\end{array}\right. \\ \nonumber
X_{z_2}(k)&=&\left\{\begin{array}{ll}
X_3 & \mbox{if } k<k_c\\
X_4 & \mbox{if } k\geq k_c.
\end{array}\right.
\eea
Here though, we have rather chosen to implement the traditional binning method with some control on the transition as: 
\bea
X_{z_1}(k) &=& \frac{X_2+X_1}{2}+\frac{X_2-X_1}{2}\tanh{\frac{k-k_c}{k_{tw}}} \label{eq:EvoBinKTrue} \\ \nonumber
X_{z_2}(k) &=& \frac{X_4+X_3}{2}+\frac{X_4-X_3}{2}\tanh{\frac{k-k_c}{k_{tw}}},
\eea
where $k_{tw}$ is the transition width between $k$ bins.  
We set $k_{tw} = k_c/10$ since we want to imitate traditional binning and ensure the transition between the bins is very rapid compared to the scale at which the transition occurs, thus the evolution maintains the appearance of being described by actual bins while having a functional definition. Although $k_{tw}$ can be used to smooth the transition between the bins but doing so to remove the appearance of actual bins also hinders the users ability to control the value of the MG parameters at $k=0$, as is easily done in the hybrid method. We leave it to the users to choose their favorite method.

\begin{figure}
\begin{center}
\begin{tabular}{|c|}
\hline
{\includegraphics[width=3.5in,height=2.8in,angle=0]{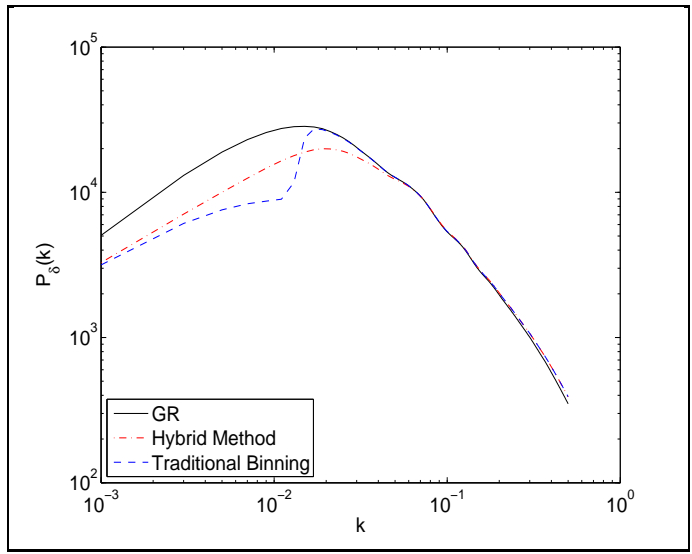}}\\
\hline  
\end{tabular}
\caption{\label{fig:PKbin}
We compare the effect the evolution of the MG parameters has on the matter power spectrum using both the binning and the new hybrid methods.  The solid black line is the matter power spectrum produced using the best fit WMAP7 parameters. The blue dashed line was produced using the traditional binning method, and the red dash-dotted line was produced using the new hybrid method.  The hybrid method produces a more realistic, smooth matter power spectrum and thus is physically well motivated.  These plots were produced using parameters $Q_1\,=\, 1.20, \, Q_2\, =\, 0.88, \, Q_3\, =\, 1.52, \,Q_4\, = \, 0.95, \mathcal{D}_1\,=\, 1.03, \, \mathcal{D}_2\, =\, 0.95, \, \mathcal{D}_3\, =\, 0.99, \,\mathcal{D}_4\, = \, 0.93,\, z_{TGR}\, =\,2, \, \mbox{and } k_c\, =\, 0.01$. The amplitudes of the two modified matter power spectra are normalized to the GR amplitude at $k \sim 7\times10^{-2}$ } 
\end{center}
\end{figure}

To further illustrate the hybrid evolution, in Fig. \ref{fig:binplot} we plot an example of the evolution of the MG parameter $Q(k,a)$ using this new hybrid form with $Q_1\,=\, 1.20, \, Q_2\, =\, 1.15, \, Q_3\, =\, 1.05, \,Q_4\, = \, 1.10, \, z_{TGR}\, =\,2, \, \mbox{and } k_c\, =\, 0.01$.  In this figure, one can see the binned aspect of the MG parameter evolution in redshift space, while the evolution in scale, $k$-space, evolves monotonically.  The advantages to this smooth evolution in scale are illustrated further in Fig. \ref{fig:PKbin}, where we compare the matter power spectrum produced using an identical set of parameters.  The hybrid method produces a much more physical and smooth matter power spectrum.    
%%
%%%%%%%%%%%
%%%%%%%%%%%%%%%%%%%%%%%%%%%%%%%%%%%%%%%%%%%%%%%%%%%%%%
\section{Modified Gravity Growth Equations and Numerical Implementations}
%%%%%%%%%%%%%%%%%%%%%%%%%%%%%%%%%%%%%%%%%%%%%%%%%%%%%%
%%%%%%%%%%%
%%

%%
%%%%%%%%%%%
%%%%%%%%%%%%%%%%%%%%%%%%%%%%%%%%%%%%%%%%%%%%%%%%%%%%%%%%%%%%%%%%%
\subsection{Synchronous Gauge Variables and CMB Implementation}
%%%%%%%%%%%%%%%%%%%%%%%%%%%%%%%%%%%%%%%%%%%%%%%%%%%%%%%%%%%%%%%%%
%%%%%%%%%%%
%%
We will first focus on our modifications to the publicly available code \texttt{CAMB} (\textbf{C}ode for \textbf{A}nisotropies in the \textbf{M}icrowave \textbf{B}ackground) \cite{LewisCAMB} which calculates the various CMB  anisotropy spectra ($C_\ell^{TT}$, $C_\ell^{TE}$, $C_\ell^{EE}$, $C_\ell^{BB}$) as well as the three-dimensional matter power spectrum $P_\delta(k,z)$.  These observables are very powerful in constraining both the growth history of structure in the Universe as well as the expansion history of the Universe, and thus are very useful in constraining  the parameters we use to test general relativity.

\texttt{CAMB} is written in the synchronous gauge. Instead of using the metric potentials $\phi$ and $\psi$ of the conformal Newtonian gauge, it uses the metric potentials $h$ and $\eta$ consistent with the notation of Ma and Bertschinger \cite{Ma}.  From Eq. (18) of \cite{Ma}, we know the metric potentials in the two gauges are related to one another by
\bea
\phi & = & \eta -\mathcal{H}\alpha,
\label{eq:PhiSG} \\
\psi &= & \dot{\alpha}+\mathcal{H} \alpha,
\label{eq:PsiSG}\\
\mbox{where } k^2\alpha & = & \frac{\dot{h}}{2} +3\dot{\eta}.
\label{eq:alphaSG}
\eea 
\texttt{CAMB} evolves the metric potential $\eta$ (actually evolving $k\eta$) as well as the matter perturbations, $\delta_i$, heat flux,$q_i$, and the shear stress $\sigma_i$ for each matter species in the synchronous gauge according to the evolution equations given in \cite{Ma}.  Additionally, the \texttt{CAMB} variables $\sigma_{CAMB}$ and $\mathcal{Z}$ are evaluated at each time step. They are useful for defining the evolution of the matter perturbations and are defined as
\bea
\sigma_{CAMB} \equiv k\alpha = \frac{k(\eta-\phi)}{\mathcal{H}},
\label{eq:sigcamb}\\
\mathcal{Z} \equiv \frac{\dot{h}}{2k} = \sigma_{CAMB}- 3\frac{\dot{\eta}}{k}.
\eea
Using these variables allows \texttt{CAMB} to be written in such a way that the evolution of all other variables is changed simply by adjusting the evolution of the metric potential $\eta$.  Thus it is important that we derive an equation for the evolution of $\eta$ consistent with the modified growth equations (\ref{eq:ModP}) and (\ref{eq:ModPSum}).  To do this, we begin by subbing (\ref{eq:PhiSG}) into (\ref{eq:ModP}) and taking the time derivative.  This gives
\be
\dot{\eta}=-\frac{1}{2k^2}\sum_i\left[\tilde{\rho_i}(a)(\dot{Q}\Delta_i +Q\dot{\Delta}_i)+Q\Delta_i\frac{d}{d\tau}\tilde{\rho_i}(a)\right]+\dot{\mathcal{H}}\alpha +\mathcal{H}\dot{\alpha}
\label{eq:etadot1}
\ee
 where $\tilde{\rho_i}(a) = 8\pi G a^2\rho_i$. Next subbing (\ref{eq:ModP}) and (\ref{eq:PsiSG})  into (\ref{eq:ModPSum}) gives an expression for $\dot{\alpha}$:
\be 
\dot{\alpha} = -\mathcal{H}\alpha - \frac{1}{2k^2}\sum_i\tilde{\rho_i}(a)\left[(2\mathcal{D}-Q)\Delta_i + 3Q(1+w_i)\sigma_i \right].
\label{eq:alphadot}
\ee
Now subbing Eqs. (\ref{eq:Deltaevo}), (\ref{eq:alphadot}), as well as the time derivative of $\rho_i$ from matter conservation into (\ref{eq:etadot1}) we get
\be
\dot{\eta}=\frac{-1}{2k^2}\sum_i\tilde{\rho_i}(a)\left[\dot{Q}\Delta_i - \mathcal{H}Q\Delta_i + 3Q(1+w_i)\left(\dot{\phi} + \mathcal{H}\psi\right) - Q f_1\frac{q_i}{k} +2\mathcal{H}\mathcal{D}\Delta_i-\mathcal{H}Q\Delta_i\right] -(\mathcal{H}^2-\dot{\mathcal{H}})\alpha,
\label{eq:etadot2}
\ee
where
\be
f_1  =  k^2 + 3 (\mathcal{H}^2-\dot{\mathcal{H}}).
\ee
Next we can sub in for $\psi$ and $\dot{\phi}$ in Eq. (\ref{eq:etadot2}) by using (\ref{eq:PsiSG}) and the time derivative of (\ref{eq:PhiSG}). This gives:
\be
\dot{\eta}=\frac{-1}{2k^2}\sum_i\tilde{\rho_i}(a)\left[\dot{Q}\Delta_i - \mathcal{H}Q\Delta_i + 3Q(1+w_i)\left(\dot{\eta} + (\mathcal{H}^2-\dot{\mathcal{H}})\alpha\right) - Q f_1\frac{q_i}{k} + 2\mathcal{H}\mathcal{D}\Delta_i - \mathcal{H}Q\Delta_i\right] -(\mathcal{H}^2-\dot{\mathcal{H}})\alpha.
\label{eq:etadot3}
\ee

We want an equation where all the variables on the right-hand-side are in the synchronous gauge, as those are the variables evolved by \texttt{CAMB}.  We already know that $\Delta$ and $\sigma$ are gauge invariant, so we need not worry about them.  However the $q_i$ in the above equations is still in the conformal Newtonian gauge.  It transforms to the synchronous gauge according to Eq. 27b of \cite{Ma} which by converting, as described above, from $\theta$ to $q$ can be written as
\be
q_i^{(c)}=q_i^{(s)}+(1+w_i)k\alpha.
\label{eq:qtransform}
\ee 
So finally subbing (\ref{eq:qtransform}) into (\ref{eq:etadot3}) and collecting the terms we have [equivalent to Eq. A8 in \cite{Bean2010}]
\be
\dot{\eta}=\frac{-1}{2f_Q}\left \{2(\mathcal{H}^2-\dot{\mathcal{H}})k^2\alpha +\sum_i\tilde{\rho_i}(a)\left[ \left(2\mathcal{H}\left[\mathcal{D}-Q\right]+\dot{Q}\right)\Delta_i -Q(1+w_i)k^2\alpha -  Q f_1\frac{q_i}{k} \right]\right\},
\label{eq:etadotfin}
\ee
with
\be
f_Q  =  k^2 + \frac{3}{2} Q\sum_i\tilde{\rho_i}(1+w_i).
\ee

After the metric potentials and matter perturbations are evolved the next major change is to redefine the derivatives of the Newtonian metric potentials, $\dot{\phi}+\dot{\psi}$, which go into evaluating the ISW effect in the CMB temperature anisotropy.  We can get this quantity quickly by applying our knowledge that the quantities $\Delta_i$ and $\sigma_i$ are invariant in transformations between the synchronous and conformal Newtonian gauges.  Thus simply taking the time derivative of (\ref{eq:ModPSum}) and subbing in for $\dot{\Delta}$ and $\dot{\tilde{\rho}}_i$ we get:
\bea
\dot{\phi}+\dot{\psi} = \frac{1}{k^2}\sum_i\tilde{\rho_i}(a)\Bigg{\{}\left[((1+3w_i)Q+2\mathcal{D})\mathcal{H}-\dot{Q}\right]\frac{3(1+w_i)\sigma_i}{2} -\frac{3Q(1+w_i)\dot{\sigma}_i}{2} \\ \nonumber
+(\mathcal{D}\mathcal{H}-\dot{\mathcal{D}})\Delta_i+\mathcal{D}(1+w_i)\left(k^2\alpha-3\dot{\eta}\right)+ \mathcal{D} f_1\frac{q_i}{k} \Bigg{\}}.
\eea
Other small changes to the scalar source term are necessary and were calculated using a slightly modified version of the Maple worksheet {\it lineofsight.txt} available from \cite{LewisCAMB}.  

Some additional modifications though still remain to alleviate problems caused by the likely possibility that bad combinations of MG parameters will be introduced to \texttt{CAMB} by the sampling routine of \texttt{CosmoMC}.  We encountered three routines which routinely ran into problems with bad combinations of MG parameters, the reionization module, the halofit module and the CMB lensing module.  To alleviate these issues we added error variables to these routines where in the event a bad parameter combination causes these modules to fail, the calculation is aborted and that parameter combination is rejected rather than causing the \texttt{CosmoMC} to stop.  With all of these changes in place \texttt{CAMB} is able to calculate and output all the modified CMB spectra as well as the modified matter power spectrum.

\subsection{Weak-Lensing Implementation}
%%%%%%%%%%%%%%%%%%%%%%%%%%%%%%%%%%%%%%%%%%%%%%%%%%%%%%%%%%
Weak-lensing tomography shear-shear cross correlations are useful to constrain both the expansion history and the growth history of structure in the Universe.  As discussed above this data can give us direct measurements of the parameter $\mathcal{D}$,;thus it is important to have a module for \texttt{CosmoMC} which takes advantage of this data.  Here we discuss our module using the weak-lensing tomography shear-shear cross correlations of the HST-COSMOS survey recently compiled by Schrabback et. al. \cite{Schrabback2010}.  {It is worth noting here that this data set probes scales on which perturbations are not described by linear theory alone.  The parametrized modified growth equations (\ref{eq:ModP}) and (\ref{eq:ModPSum}) describe deviations from GR on linear scales only.  Though we use the halofit module to describe the nonlinear part of the power-spectrum, it should be noted that this may not be completely accurate and should be considered a source of possible systematic error for our MG parameter constraints.}  The authors of \cite{Schrabback2010} perform a refined analysis of the HST-COSMOS survey of \cite{Scoville2007}, in combination with the COSMOS-30 photometric redshift catalog provided by \cite{Ilbert2009}.  The shear-shear cross correlations were calculated between 6 redshift bins, $0.0<z<0.6$, $0.6<z<1.0$, $1.0<z<1.3$, $1.3<z<2.0$, $2.0<z<4.0$ and a sixth bin that contains all faint galaxies with a numerically estimated redshift distribution from $0.0<z<5.0$; see Fig. 6 of \cite{Schrabback2010}.  Certain exclusions were made as described in \cite{Schrabback2010}, such as the luminous red galaxies (LRGs) to avoid G-I intrinsic alignment bias and the lowest angular theta bin because of model uncertainties.  Only bright galaxies ($i<24$) were used in the first bin $(z<0.6)$. Also, autocorrelations were not used in bins $1-5$ to reduce the effect of I-I intrinsic alignments.  We modified the publicly available code for the COSMOS 3D weak-lensing built by Lesgourgues et. al. \cite{Lesgourgues} to incorporate the shear cross correlations as seen in, for example, \cite{Schrabback2010} and the MG parameters.

The shear cross correlation functions $ \xi^{kl}_{+,-}(\theta) $ between bins $k,l$ are given by
\be 
 \xi^{kl}_{+,-}(\theta) = \frac{1}{2\pi} \int^{\infty}_{0} d\ell\ \ell\  J_{0,4}(\ell\theta)P^{kl}_{\kappa} (\ell),
 \label{eq:ShearCrossCorrelations}
 \ee
 where $J_n$ is the $n^{th}$-order Bessel function of the first kind, $\ell$ is the modulus of the two-dimensional wave vector, and $P^{kl}_{\kappa}$ is the convergence cross-power spectra between bins $k,l$.  
 
In order to evaluate the convergence cross-power spectra, let us first recall that working with the metric (\ref{eq:FLRW}),  the convergence is defined as
\be
\kappa(\hat{n}) = \int^{\chi_h}_0 d\chi \, g(\chi) \, f_K(\chi) \, \frac{1}{2}\nabla^2\Big{[}\phi(f_K(\chi)\hat{n},\chi)+\psi(f_K(\chi)\hat{n},\chi)\Big{]},
\label{eq:Convergence}
\ee
with comoving radial distance $\chi$, comoving distance to the horizon $\chi_h$,comoving angular diameter distance $f_K(\chi)$, and the weighted geometric lens-efficiency factor $g(\chi)$ given by
\be 
g(\chi)  \equiv \int^{\chi_h}_{\chi} d\chi' p(\chi') \frac{f_K(\chi'-\chi)}{f_K(\chi')},
\label{eq:LensEfficiency}
\ee
corresponding to the galaxy redshift distributions $p$.  Applying the Limber approximation to (\ref{eq:Convergence}) to get the convergence cross power spectrum in the absence of any modifications of growth gives us the usual result in terms of the three-dimensional (nonlinear) power spectrum $P_{\delta}$:
\be 
P^{kl}_{\kappa}(\ell) = \frac{9H^4_0 \Omega_m^2}{4c^4} \int^{\chi_h}_{0}  d\chi \frac{g_k(\chi)g_l(\chi)}{a^2(\chi)}\, P_{\delta} \Big(\frac{\ell}{f_K(\chi)},\chi \Big),
\label{eq:3DPowerSpectrum}
\ee
 where $g_k(\chi)$ is the weighted geometric lens-efficiency factor as defined above corresponding to the $k^{th}$ redshift bin.

Now to incorporate our MG parametrization we substitute the late-time form (ie. negligible anisotropic stress) of (\ref{eq:ModPSum}) into (\ref{eq:Convergence}) and again apply the Limber approximation.  This gives us a modified convergence cross power spectrum that takes into account our MG parametrization:
\be 
P^{kl}_{\kappa}(\ell) = \frac{9H^4_0 \Omega_m^2}{4c^4} \int^{\chi_h}_{0}  d\chi \frac{g_k(\chi)g_l(\chi)}{a^2(\chi)}\, \mathcal{D}^2\Big(\frac{\ell}{f_K(\chi)},a(\chi) \Big)\, P_{\delta} \Big(\frac{\ell}{f_K(\chi)},\chi \Big),
\label{eq:3DPowerSpectrumMG}
\ee

In calculating our cross correlation functions we follow instructions from \cite{Schrabback2010} and weight the predictions for the $k^{th}$ $\theta$ bin, $\theta_{k}$, with logarithmic spaced upper and lower limits $\theta_{k,max}$ and $\theta_{k,min}$, respectively,  based on the number of galaxy pairs, $N$ for a given $\theta$ as
 \be 
 \xi(\theta_{k})= \frac{\int^{\theta_{k,max}}_{\theta_{k,min}} N(\theta') \xi_{+,-}(\theta') d\theta'}{\int^{\theta_{k,max}}_{\theta_{k,min}} N(\theta') d\theta'},
 \label{eq:WeightTheta}
 \ee
 with $N(\theta) \propto \theta(0.0004664+\theta(0.0044118-8.90878\times 10^{-5}\theta))$ \cite{Schrabback2010,Schrabback2010private_comm}.

To evaluate the likelihood we use the $160 \times 160$ dimensional covariance matrix for the corresponding weak shear correlations provided by Schrabback et. al.  \cite{Schrabback2010} applying the correction to the inverse covariance, $C^{-1}$, seen in \cite{Schrabback2010, Hartlap2007}, as
\be 
\stackrel{\ast}{C}{}^{-1} = 0.4390 \,C^{-1}.
\label{eq:InMatrix}
\ee
 for $288$ independent realizations and a $160$ dimensional data vector.  {We then account for the $10\%$ uncertainty in the numerical estimate of the galaxy redshift distribution of bin six, $p_6(z)$, by marginalizing over the nuisance parameter $f_z$ which is applied following \cite{Schrabback2010} where for the sixth bin we instead use the galaxy redshift distribution $p_6'(z,f_z)$ given by $p_6'(z,f_z) \equiv p_6(f_z z)$}.  Now the likelihood ($\mathcal{L}$) is given by:
\be
-2 \ln \mathcal{L} = \sum_{ij}\left(\xi_{th}-\xi_{obs}\right)_i \left[\stackrel{\ast}{C}{}^{-1}\right]_{i j}\left(\xi_{th}-\xi_{obs}\right)_j + \left(\frac{f_z-1}{0.1}\right)^2.
\label{eq:WLLike}
\ee

%%%
%%%%%%%%%%%%%%%%%%%%%%%%%%%%%%%%%%%%%%%%%
\subsection{ISW-Galaxy Cross Correlations Implementation}
%%%%%%%%%%%%%%%%%%%%%%%%%%%%%%%%%%%%%%%%%
%%%
As discussed in, for example, \cite{Bean2010,GongBo2010,DL2010}, cross correlations between the ISW effect and the galaxy density can be very useful in constraining the MG parameters.  Thus, we modified the publicly available module by Ho et al.\cite{ISWHo,ISWHirata} for calculating these cross correlations in the presence of a modification to gravity described by Eqs. (\ref{eq:ModP}) and (\ref{eq:ModPSum}).  Let us review some of the details of calculating these cross correlations.  

First recall, when working with the perturbed FLRW metric (\ref{eq:FLRW}) that the CMB temperature anisotropy due to the ISW effect can be written as \cite{ISWHo}
\be
\Delta T_{ISW}(\hat{n}) = \int_{\tau_r}^{\tau_0} d\tau \frac{\partial}{\partial\tau}(\phi+\psi),
\label{eq:ISW}
\ee
where $\tau_r$ is the conformal time of recombination and $\tau_0$ is the conformal time today.  The angular galaxy density fluctuations that we want to cross correlate these temperature fluctuations with are
\be
g(\hat{n}) = \int dz \, b(z) \, \Pi(z) \, \delta_m\left(f_K(\chi(z))\hat{n},z\right),
\label{eq:galdens}
\ee
where $b(z)$ is the galaxy bias, $\Pi(z)$ is the normalized selection function, and $f_K(\chi(z))$ is the comoving angular diameter distance redshift $z$.

Following \cite{ISWHo}, we can once again use the Limber approximation to write the cross-power spectrum between these two fields in the absence of any modifications to the growth as
\be
C_\ell^{gT} = \frac{3\Omega_m H^2_0 T_{CMB}}{c^2(\ell +1/2)^2}\int dz b(z)\Pi(z) \frac{H(z)}{c}D(z)\frac{d}{dz} \left[(1+z)D(z)\right]P\left(\frac{\ell+1/2}{f_K(\chi(z))}\right),
\label{eq:gTcross}
\ee
where the growth factor $D(z) = D(0) \delta_m(k,z)/\delta_m(k,0)$ [please note that this $D(z)$ is not the parameter $\mathcal{D}$ we use to test GR]  is calculated using the standard differential equation for $\delta_m(z)$ at linear scales, $\ddot{\delta}_m+2H\dot{\delta}_m+k^2\psi/a^2 = 0$ with $\psi$ given in Eq. (\ref{eq:PhiGR}).
This result was attained by considering at late times in GR the potentials in (\ref{eq:ISW}) can be written combining to (\ref{eq:P}) and (\ref{eq:2E}) as
\be
\phi(k,z) \,= \,  \psi(k,z) \,  =\, -\frac{3H_0^2}{2}(1+z)\Omega_m\frac{\delta_m(k,z)}{k^2}. 
\label{eq:PhiGR}
\ee 

When testing general relativity though, Eq. (\ref{eq:PhiGR}) no longer holds and we must instead, as with the weak-lensing cross correlations, refer to the late-time version of (\ref{eq:ModPSum}) giving
\be
\phi(k,z) +\psi(k,z) =   -3H_0^2\, \mathcal{D}(k,z)\, (1+z)\, \Omega_m\frac{\delta_m(k,z)}{k^2}.
\label{eq:PhiMG}
\ee
With these changes the ISW-galaxy cross power spectrum is given by:
\be
C_\ell^{gT} = \frac{3\Omega_m H^2_0 T_{CMB}}{c^2(\ell +1/2)^2}\int dz b(z)\Pi(z) \frac{H(z)}{c}D(z)\frac{d}{dz} \left[\mathcal{D}\left(\frac{\ell+1/2}{f_K(\chi(z))},z\right)\,(1+z)D(z)\right]P\left(\frac{\ell+1/2}{f_K(\chi(z))}\right).
\label{eq:gTcrossMG}
\ee
Additionally, to fully account for the gravity modifications, the differential equation used to calculate the $\delta(k,z)$ in $D(z)$ must be modified.  Combining Eqs. (\ref{eq:ModP}) and (\ref{eq:ModPSum}) to get $k^2\psi$, the late-time form of differential equation for $\delta(k,z)$ now reads
\be
\ddot{\delta}_m(k,z)+2H\dot{\delta}_m(k,z)-\frac{3H_0^2}{2}\left[2\mathcal{D}(k,z)-Q(k,z)\right](1+z)^3\Omega_m\delta_m(k,z) = 0.
\label{eq:deltaMG}
\ee

In order to correctly perform these cross correlations, the code by Ho et al. also corrects factor $b(z)\Pi(z)$ to account for a magnification bias on  SDSS quasars and NVSS radio sources \cite{ISWHo}. In calculating this magnification bias, the weak-lensing convergence must be considered on scales smaller than the Hubble distance \cite{Matsubara}.  Thus, we modify the code to account for changes in the weak-lensing convergence according to (\ref{eq:Convergence}) and (\ref{eq:ModPSum}) but consider only large-$k$ values of the MG parameters for these corrections in calculating the magnification bias.  

The original code by Ho et al. also has a likelihood code which uses cross correlations between weak-lensing of the CMB and the galaxy density.  A detailed description of this part of the code is contained in \cite{ISWHirata}.  We modify this portion of the code to account for the changes to the convergence in the presence of modifications to the growth as described by Eqs. (\ref{eq:ModPSum}) and (\ref{eq:Convergence}).  

%%%
%%%%%%%%%%%%%%%%%%%%%%%%%%%%%%%%%%%%%%%%%
\subsection{WiggleZ Baryon Acoustic Oscillation Likelihood}
%%%%%%%%%%%%%%%%%%%%%%%%%%%%%%%%%%%%%%%%%
%%%
Recently, the WiggleZ Dark Energy Survey released its full data set of BAO measurements \cite{BlakeBAO}.  We include a likelihood module for this data set in \texttt{ISiTGR}.  Let us quickly review the likelihood calculation for this data.

The data set released by \cite{BlakeBAO} includes measurements of the acoustic parameter $A(z)$ at three different effective redshifts:  $z=0.44$, $z=0.6$, and $z=0.73$.  The acoustic parameter was first introduced by \cite{Eisenstein2005} and is defined as
\be
A(z) = \frac{100 D_V(z) \sqrt{\Omega_m h^2}}{c\, z},
\ee
where $\Omega_m h^2$ is the physical matter density. $D_V(z)$ is the effective distance defined by:
\be
D_V(z) = \left( D^2_A(z)(1+z)^2\frac{c\, z}{H(z)}\right)^{1/3},
\ee
with the angular diameter distance to the redshift $z$, $D_A(z)$.

To calculate the likelihood, the theoretical value of $A(z)$ is calculated at each of the effective redshifts.  Then using the covariance matrix for the data given in \cite{BlakeBAO}, the likelihood for a given model is expressed as
\be
-2 \ln \mathcal{L} = \sum_{ij}\left(A_{th}-A_{obs}\right)_i \left[C^{-1}\right]_{i j}\left(A_{th}-A_{obs}\right)_j.
\ee
%%%
%
\begin{figure}[t]
\begin{center}
\begin{tabular}{|c|}
\hline 
{\includegraphics[width=3.3in,height=1.35in,angle=0]{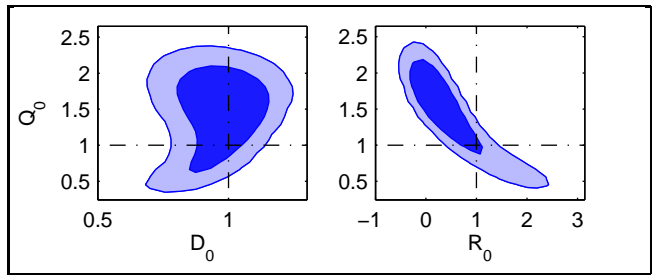}} \\ \hline
\end{tabular}
\caption{\label{figure:FUNC}
We plot the $68\%$ and $95\%$ C.L. constraints on the MG parameters $Q_0$, $R_0$, and $\mathcal{D}_0$ from evolving the parameters with a functional form. We use all available data sets included in \texttt{ISiTGR}: SN, BAO, AGE, $H_0$, CMB, MPK, ISW, and WL.  All constraints using this method are fully consistent with GR, however, as we show further, the correlations between these parameters and some cosmological parameters are significant.} 
\end{center}
\end{figure}
\begin{figure}[t]
\begin{center}
\begin{tabular}{|c||c|}
\hline 
{\includegraphics[width=3.3in,height=1.35in,angle=0]{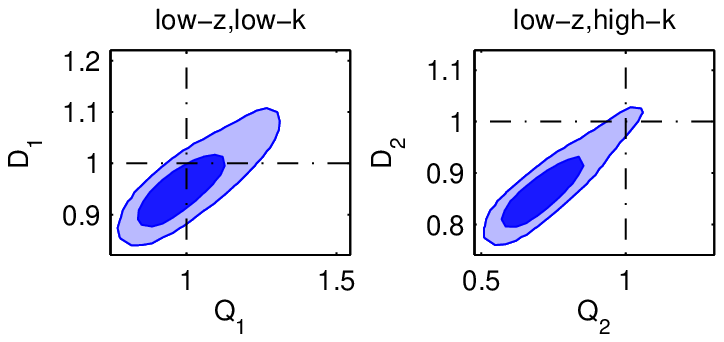}}& {\includegraphics[width=3.3in,height=1.35in,angle=0]{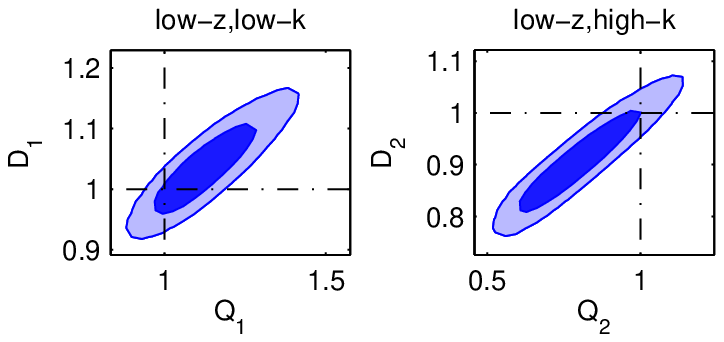}} \\ \hline
{\includegraphics[width=3.3in,height=1.35in,angle=0]{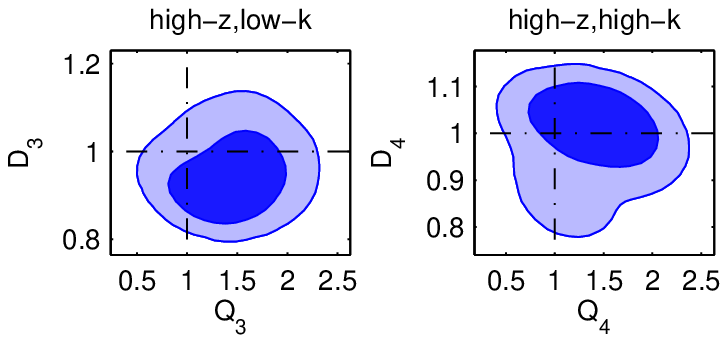}}&
{\includegraphics[width=3.3in,height=1.35in,angle=0]{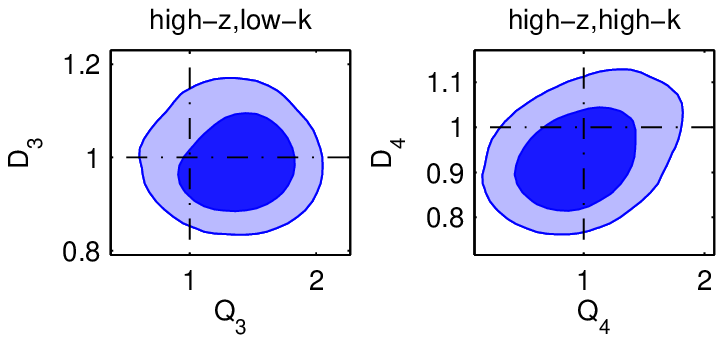}} \\
\hline
\end{tabular}
\caption{\label{figure:Bin}
Left panel: We plot the $68\%$ and $95\%$ C.L. constraints on the MG parameters $Q_i$ and $\mathcal{D}_i$, $i=1,2,3,4$ from using traditional bins for $k$ and $z$.   $z$-bins are $0<z\le1$ and $1<z\le 2$ with GR assumed for $z>2$ and $k$-bins are $k\le0.01$ and $k>0.01$. These constraints come from using all available data sets included in \texttt{ISiTGR}: SN, BAO, AGE, $H_0$, CMB, MPK, ISW, and WL. While all constraints are consistent with GR, bin 2 seems to be indicating some tensions with GR values just within the $95\%$ C.L.
Right panel: We plot the $68\%$ and $95\%$ C.L. constraints on the MG parameters $Q_i$ and $\mathcal{D}_i$, $i=1,2,3,4$ from using the new hybrid binning method.  $z$-bins are still $0<z\le1$ and $1<z\le 2$ with GR assumed for $z>2$, while the transition scale for  $k$ evolution is $k_c=0.01$. Again, we use all available data sets included in \texttt{ISiTGR}: SN, BAO, AGE, $H_0$, CMB, MPK, ISW, and WL.  All constraints using this method are fully consistent with GR but as we show further, the correlations between these parameters and some cosmological parameters are significant.} 
\end{center}
\end{figure}
\begin{center}
\begin{table}[t]
\begin{tabular}{|c|c|c|c|c|c|c|c|c|}\hline
\multicolumn{9}{|c|}{{\large \bfseries Correlation coefficients between $\Omega_m$ and $\sigma_8$ and the MG parameters}}\\ \hline
\multicolumn{9}{|c|}{{\large MG parameters evolved using traditional binning}}\\ \hline
Parameter&$Q_1$&$Q_2$&$Q_3$&$Q_4$&$\mathcal{D}_1$&$\mathcal{D}_2$& $\mathcal{D}_3$&$\mathcal{D}_4$\\ \hline
$\Omega_m$&\,0.0116 \,& \, -0.0535 \,  & \,  -0.0338  \, &  \, -0.0371 \,  &  \, 0.0114 \,  & \,  -0.0988 \,  & \,  0.0190  \, & \,  -0.0818 \,  \\ \hline
$\sigma_8$&  0.0937  &0.1540&-0.1583&-0.4267&0.2093&0.2857&0.2377&0.0185\\  \hline
\multicolumn{9}{|c|}{}\\ \hline
\multicolumn{9}{|c|}{{\large MG parameters evolved using hybrid binning}}\\ \hline
Parameter&$Q_1$&$Q_2$&$Q_3$&$Q_4$&$\mathcal{D}_1$&$\mathcal{D}_2$& $\mathcal{D}_3$&$\mathcal{D}_4$\\ \hline
$\Omega_m$&0.1571&-0.1639&-0.003&-0.0105&0.1814&-0.2172&-0.0056&-0.0070\\ \hline
$\sigma_8$&0.1692&0.4282&-0.1076&0.1676&0.2147&0.5390&0.1937&0.0716\\ \hline
\end{tabular}
\\
\begin{tabular}{|c|c|c|c|}\hline 
\multicolumn{4}{|c|}{}\\ \hline
\multicolumn{4}{|c|}{{\large MG parameters evolved using the functional form}}\\ \hline
Parameter&$Q_0$&$\mathcal{D}_0$&$R_0$\\ \hline
$\Omega_m$& \,\,\, \,\,\,-0.1173 \,\,\, \,\,\, & \,\,\, \,\,\,-0.0294 \,\,\, \,\,\,& \,\,\, \,\,\,0.0542 \,\,\, \,\,\, \\ \hline
$\sigma_8$&-0.5409&-0.2631&0.5742\\ \hline
\end{tabular}
\caption{\label{table:Corr}
We list the correlation coefficients for $\Omega_m$ and $\sigma_8$ and the various MG parameters.  We see that the MG parameters for all evolution methods are significantly more correlated with $\sigma_8$ than $\Omega_m$, although some correlation with $\Omega_m$ does exist.  While traditional binning shows overall the least amount of correlation between the MG parameters and the two cosmological parameters, it, as shown in Fig. \ref{fig:PKbin}, has a problematic jump in the matter power spectrum.  The hybrid binning is the next least correlated of the methods, with parameters from the functional form evolution showing the most amount of correlation with the two cosmological parameters compared here.
}
\end{table}
\end{center} 
\begin{figure}[t]
\begin{center}
\begin{tabular}{|c|}
\hline 
{\includegraphics[width=3.25in,height=2.5in,angle=0]{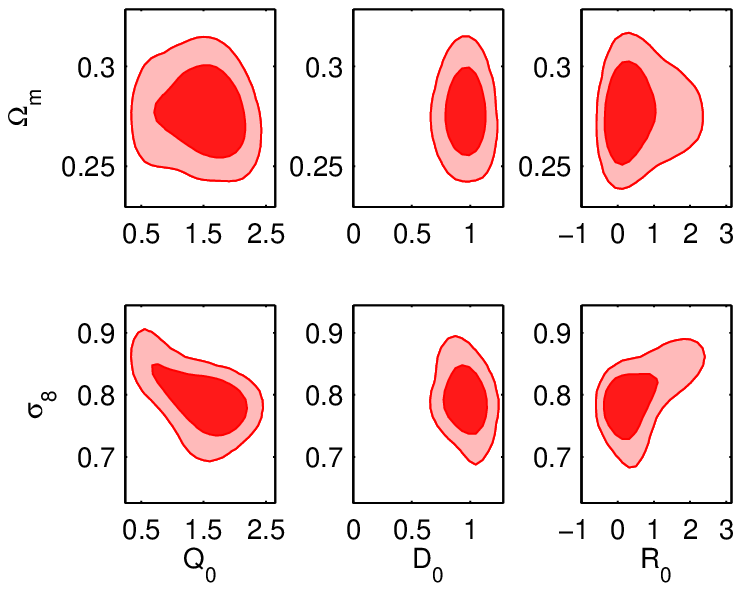}} \\
\hline   
\end{tabular}
\caption{\label{figure:CorrGRIDR}
We plot here the 2D confidence contours for $\Omega_m$ and $\sigma_8$ and the MG parameters $Q_0$ and $\mathcal{D}_0$, from using the functional form to evolve the MG parameters.  As seen in Table \ref{table:Corr} this evolution method overall has the most amount of correlations between the MG parameters and the two cosmological parameters.} 
\end{center}
\end{figure}
\begin{figure}[b]
\begin{center}
\begin{tabular}{|c|}
\hline 
{\includegraphics[width=4.33in,height=2.5in,angle=0]{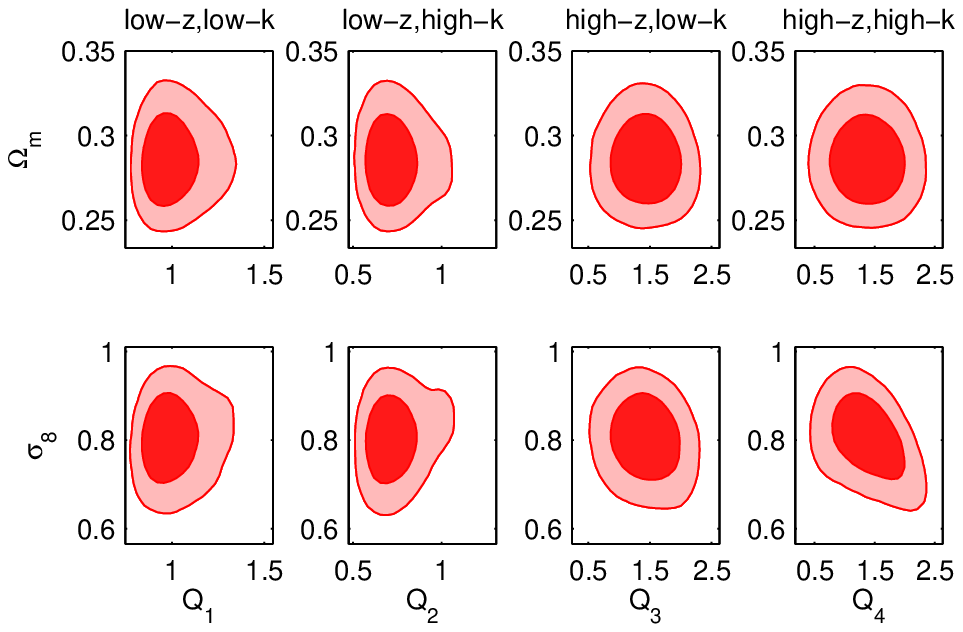}} \\ \hline
{\includegraphics[width=4.33in,height=2.5in,angle=0]{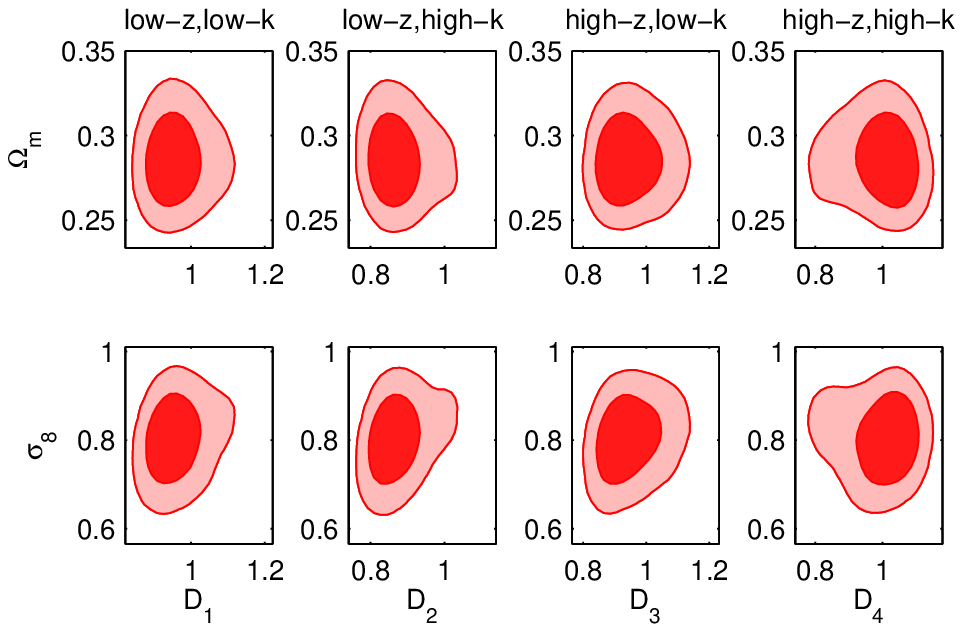}} \\
\hline
\end{tabular}
\caption{\label{figure:CorrGRIDB}
We plot here the 2D confidence contours for $\Omega_m$ and $\sigma_8$ and the MG parameters $Q_i$ and $\mathcal{D}_i$, $i=1,2,3,4$ from using traditional bins for $k$ and $z$.  As seen in Table \ref{table:Corr} this evolution method has the least amount of correlation between the MG parameters and the two cosmological parameters of the three evolution methods, but as discussed earlier suffers from the appearance of a jump in the matter power spectrum due to the rapid transition between scale bins.} 
\end{center}
\end{figure}   
\begin{figure}[t]
\begin{center}
\begin{tabular}{|c|}
\hline 
{\includegraphics[width=4.33in,height=2.5in,angle=0]{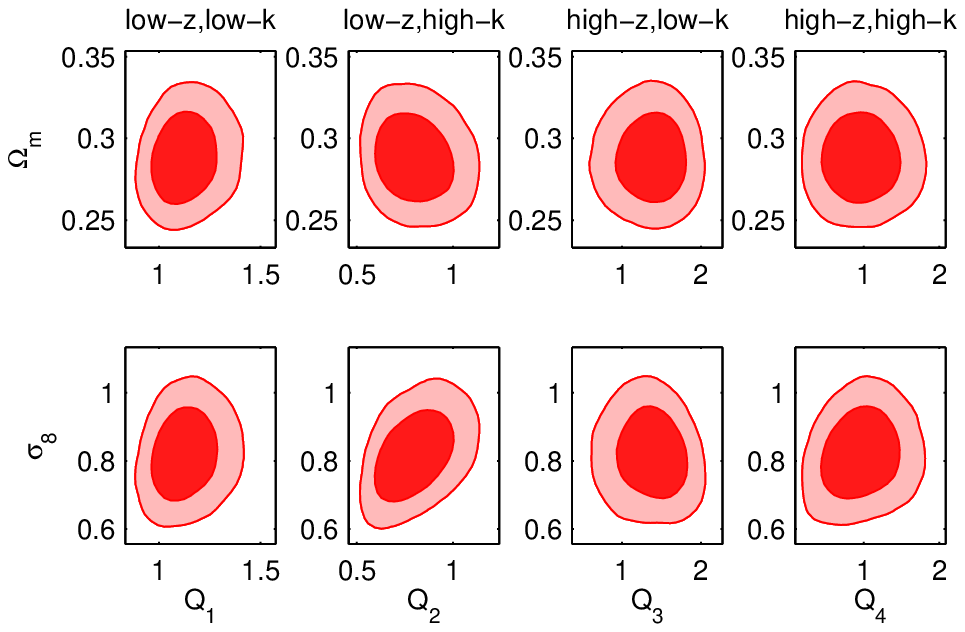}}\\ \hline
{\includegraphics[width=4.33in,height=2.5in,angle=0]{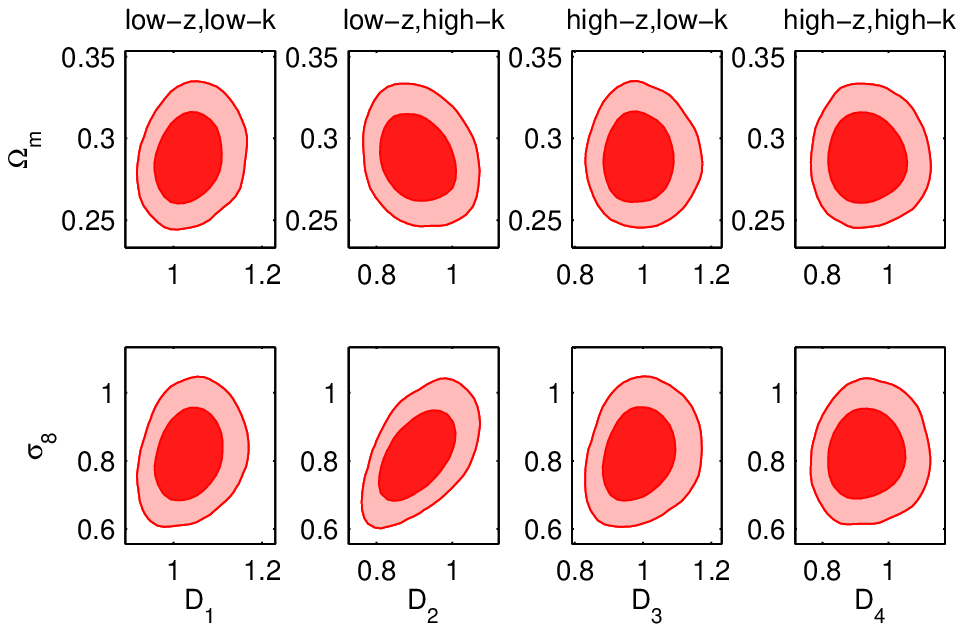}} \\
\hline
\end{tabular}
\caption{\label{figure:CorrGRIDH}
We plot here the 2D confidence contours for $\Omega_m$ and $\sigma_8$ and the MG parameters $Q_i$ and $\mathcal{D}_i$, $i=1,2,3,4$ from using the new hybrid method to evolve the MG parameters.  As seen in Table \ref{table:Corr} this evolution method has only a moderate amount of correlation between the MG parameters and the two cosmological parameters.} 
\end{center}
\end{figure}
\section{Correlations obtained between cosmological parameters and modified gravity growth parameters}
We provide here our results about various constraints and correlation coefficients found between core cosmological parameters and modified growth parameters. We use the standard definition for the correlation coefficient:
\be
Corr(p_x,p_y)=\frac{Cov(p_x,p_y)}{\sigma({p_x})\sigma({p_y})}
\ee
where $p_x$, $p_y$ are the parameters, $Cov(p_x,p_y)$ is the covariance of the two parameters, and $\sigma({p_x})$ and $\sigma({p_y})$ are their respective standard deviations. 

In addition to varying a given set of modified gravity parameters, we vary the six core cosmological parameters:  $\Omega_bh^2$ and the $\Omega_c h^2$, the baryon and cold-dark matter physical density parameters, respectively; $\theta$, the ratio of the sound horizon to the angular diameter distance of the surface of last scattering; $\tau_{rei}$, the reionization optical depth; $n_s$, the spectral index; and $\ln10^{10} A_s$, the amplitude of the primordial power spectrum.  Additionally, the results illustrated here always use the following data sets for expansion history constraints, WiggleZ BAO measurements \cite{BlakeBAO}, the supernovae Union2 compilation of the Supernovae Cosmology Project (SCP)  \cite{Union2} and references of other compiled supernovae (SN) therein.  We also use the prior on $H_0=74.2\pm 3.6$ km/s/Mpc given by \cite{riess}, and a prior on the age of the Universe (AGE) $10\,$Gyrs$<$AGE$<20\,$Gyrs.

We include in our analysis the three parametrization approaches: functional form, binned method, and hybrid method. 

First, we utilize the functional form to evolve the MG parameters.  For simplicity, we choose to assume the modifications are scale-independent ($k_c=\infty$) and evolve the parameters $Q$ and $R$ with the functional forms described by (\ref{eq:BeanEvo}):
\bea
Q(k,a) &=& \left(Q_0-1\right)a^s +1, 
\label{eq:QEvoEx}\\
R(k,a) &=& \left(R_0 -1\right)a^s +1.
\label{eq:REvoEx}
\eea
So, in addition to varying the six core cosmological parameters listed above, we vary the MG parameters $Q_0$, $\mathcal{D}_0$ (inferring $R_0$ from $Q_0$ and $\mathcal{D}_0$),  and $s$.  To constrain these parameters we use all of the available data sets: the WMAP7 temperature and polarization spectra (CMB) \cite{WMAP7all}, the matter power spectrum (MPK) from the Sloan Digital Sky Survey (SDSS) DR7 \cite{BAOReid}, the ISW-galaxy cross correlations \cite{ISWHo,ISWHirata}, and the refined HST-COSMOS weak-lensing tomography \cite{Schrabback2010}.  

Second, we bin the MG parameters rather than evolve with some functional form.  We bin traditionally in both $z$ and $k$ as described by (\ref{eq:EvoBinZ}) and (\ref{eq:EvoBinKTrue}) with $z_{div} = 1.0$, $z_{TGR}=2.0$, $z_{tw} =0.05$, $k_c= 0.01$ and $k_{tw}=0.001$:
\bea
Q(k,a) &=&\frac{1 + Q_{z_1}(k)}{2}+\frac{Q_{z_2}(k) - Q_{z_1}(k)}{2}\tanh{\frac{z-1}{0.05}}+\frac{1 - Q_{z_2}(k)}{2}\tanh{\frac{z-2}{0.05}},\label{eq:EvoBinZEx}\\ \nonumber
\mathcal{D}(k,a) &=&\frac{1 + \mathcal{D}_{z
_1}(k)}{2}+\frac{\mathcal{D}_{z_2}(k) - \mathcal{D}_{z_1}(k)}{2}\tanh{\frac{z-1}{0.05}}+\frac{1 - \mathcal{D}_{z_2}(k)}{2}\tanh{\frac{z-2}{0.05}},
\eea
with
\bea
Q_{z_1}(k) &=& \frac{Q_2+Q_1}{2}+\frac{Q_2-Q_1}{2}\tanh{\frac{k-0.01}{0.001}},
\label{eq:QEvoBinKTrueEx} \\ \nonumber
Q_{z_2}(k) &=& \frac{Q_4+Q_3}{2}+\frac{Q_4-Q_3}{2}\tanh{\frac{k-0.01}{0.001}},\\
\mathcal{D}_{z_1}(k) &=& \frac{\mathcal{D}_2+\mathcal{D}_1}{2}+\frac{\mathcal{D}_2-\mathcal{D}_1}{2}\tanh{\frac{k-0.01}{0.001}},
\label{eq:DEvoBinKTrueEx} \\ \nonumber
\mathcal{D}_{z_2}(k) &=& \frac{\mathcal{D}_4+\mathcal{D}_3}{2}+\frac{\mathcal{D}_4-\mathcal{D}_3}{2}\tanh{\frac{k-0.01}{0.001}}.
\eea
So, the $z$-bins are $0<z\le1$ and $1<z\le 2$ and GR is assumed for $z>2$, while the $k$-bins are $k\le0.01$ and $k>0.01$. We vary all eight MG parameters, $\mathcal{D}_i\, ,Q_i\,\,\,i=1,2,3,4$, in addition to the six core cosmological parameters and again we use all of the available data sets: CMB, MPK, ISW, and WL.  

Finally, we evolve the MG parameters using the newly introduced hybrid method with evolution with redshift ($z$) dependence identical to that of traditional binning method, Eq. (\ref{eq:EvoBinZEx}), and scale ($k$) dependence described by Eq. (\ref{eq:EvoBinKExp}) with $k_c= 0.01$:
\bea
Q_{z_1}(k) &=& Q_1 e^{-k/0.01}+Q_2(1-e^{-k/0.01}), 
\label{eq:QEvoBinKExpEx} \\ \nonumber
Q_{z_2}(k) &=& Q_3 e^{-k/0.01}+Q_4(1-e^{-k/0.01}),\\
\mathcal{D}_{z_1}(k) &=& \mathcal{D}_1 e^{-k/0.01}+\mathcal{D}_2(1-e^{-k/0.01}),
\label{eq:DEvoBinKExpEx} \\ \nonumber
\mathcal{D}_{z_2}(k) &=& \mathcal{D}_3 e^{-k/0.01}+\mathcal{D}_4(1-e^{-k/0.01}).
\eea
So, just as with traditional binning, the $z$-bins are $0<z\le1$ and $1<z\le 2$ and GR is assumed for $z>2$ while the scale dependence transitions exponentially between two values for each redshift bin with a decay constant $k_c=0.01$. We again vary all eight MG parameters, $\mathcal{D}_i\, ,Q_i\,\,\,i=1,2,3,4$, in addition to the six core cosmological parameters and use all of the available data sets: CMB, MPK, ISW, and WL.  

We plot the $68\%$ and $95\%$ 2D contours for the MG parameters in Figs. \ref{figure:FUNC} and \ref{figure:Bin} for the three methods respectively. We find that the constraints on modified gravity parameters are consistent with general relativity for all methods; however, the correlations between these parameters and some cosmological parameters are significant as we describe below. 
 
In Table \ref{table:Corr}, we report the results obtained for the correlation coefficients between the modified gravity parameters and cosmological parameters.  The table shows that modified gravity parameters are significantly correlated with $\sigma_8$ and mildly correlated with $\Omega_m$. 
For example, the degeneracies between $\sigma_8$ and modified gravity parameters are found to be substantial, especially for the functional form evolution method where $Corr(Q_0,\sigma_8)=-0.54$, $Corr(R_0,\sigma_8)=0.57$, and $Corr(\mathcal{D}_0,\sigma_8)=-0.26$. 
For the hybrid method, about half of the bins seem to present some strong correlations, with the strongest being $Corr(\mathcal{D}_2,\sigma_8)=0.54$. The traditional  binning method seems to present less correlation than the  hybrid method, but again about half of the bins have strong correlations with the largest being $Corr(Q_4,\sigma_8)=-0.43$. 

Correlations between MG parameters and $\sigma_8$ have been discussed briefly by \cite{DL2010}.  They too found that the correlations were significant.  We find that for both the hybrid and traditional binning method that correlations between $\sigma_8$ and the $\mathcal{D}$ parameters are positive meaning higher values of $\mathcal{D}$ gives rise to higher values of $\sigma_8$.  We also find that, with the exception of the parameter $\mathcal{D}_4$, the high $k$ parameters are more strongly correlated than the low $k$ parameters, agreeing with the conclusions of \cite{DL2010} that these correlations are strongly dependent on the matter power spectrum, weak lensing, and ISW data sets.  The corresponding plots are given in Figs. \ref{figure:CorrGRIDR}, \ref{figure:CorrGRIDB}, and \ref{figure:CorrGRIDH}, for the functional form method, traditional binning method, and hybrid method respectively.  Looking at Fig. \ref{figure:CorrGRIDB} we can see that there is not much difference between the slope of contours for $\sigma_8$ and $\mathcal{D}_2$ or $\mathcal{D}_4$, while the correlations coefficients given in Table \ref{table:Corr} are much larger for $\mathcal{D}_2$ this can be explained by the bulges in each of those contours.  For $\mathcal{D}_2$ this bulge is toward the higher values of that parameter and at higher values of $\sigma_8$ thereby increasing the positive correlation, while for $\mathcal{D}_4$ this bulge prefers lower values of said parameter, but still higher values of $\sigma_8$ presenting an anticorrelation that washes out the underling positive correlation of the rest of the parameter space.  These bulges are caused by parameter constraints that are not completely Gaussian, as also seen in Fig. \ref{figure:Bin}.  Non-Gaussianity in the constraints also accounts for the odd shapes of the contours in Fig. \ref{figure:CorrGRIDR}, showing the correlations for the functional form MG parameters.

%%%%%%%%%%%%%%%%%%%%%%%%%%%%%%%%%%%%%%%%%
\section{Conclusion}
%%%%%%%%%%%%%%%%%%%%%%%%%%%%%%%%%%%%%%%%%
Testing general relativity on cosmological scales is a topic of great interest in cosmology today.  It allows an exploration of the origin of cosmic acceleration, as well as testing possible extensions to general relativity. Future high precision data will allow us to test gravity on cosmological scales to a very high level of precision. 
In this work we focused on the presence of correlations between 
modified gravity growth parameters and core cosmological parameters finding that some of them can be very strong. 
We provided the underlying modified growth equations and their evolution. We implemented the time and scale dependencies of the MG parameters using a functional form method, a binning method, as well as, a new hybrid method. The hybrid method provides a smooth evolution in scale and a binned redshift (time) dependence that was shown to be more robust than time functional forms.
We used the most recent available cosmological data sets including BAO measurements from the WiggleZ Dark Energy Survey \cite{BlakeBAO} as well as the Two-Degree Field and DR7 Sloan Digital Sky surveys \cite{BAOReid,Percival2009}; priors on $H_0$ from \cite{riess};  the Wilkinson Microwave Anisotropy Probe (WMAP) 7-year CMB temperature and polarization spectra \cite{WMAP7all};  the supernovae Union2 compilation, which includes the 557 type-Ia SCP \cite{Union2} and references of other compiled supernovae therein; the MPK from the SDSS DR7 \cite{BAOReid}; the ISW-galaxy cross correlations using the 2MASS and SDSS LRG galaxy surveys \cite{ISWHo,ISWHirata}; and the recently refined HST-COSMOS weak-lensing tomography analysis in \cite{Schrabback2010}. 
We also described our implementation of the modified growth equations into a
numerical framework, \texttt{ISiTGR}, an integrated set of modified modules for the packages \texttt{CosmoMC} \cite{cosmomc} and \texttt{CAMB} \cite{LewisCAMB} to be used to this end. \texttt{ISiTGR} modifies multiple publicly available codes:  \texttt{CAMB} \cite{LewisCAMB}, \texttt{CosmoMC} \cite{cosmomc},  the ISW-galaxy cross correlation code \cite{ISWHo,ISWHirata}, and the original COSMOS weak-lensing code \cite{Lesgourgues}. When using and referring to this package, the original codes should also be cited, as well as the above data sets when used in a particular analysis.  
Our analysis found that strong correlations between $\sigma_8$ and 
modified gravity parameters are present, in addition to mild correlations with other cosmological parameters such as $\Omega_m$. The degeneracies between $\sigma_8$ and modified gravity parameters are substantial for the functional form and also for nearly half of the bins in the hybrid and binned methods. These degeneracies will need to be addressed when using future high precision data to perform these tests. 
\acknowledgements
{We thank T. Schrabback for providing the refined HST COSMOS data. We thank A. Lewis for useful comments. M.I. acknowledges that this material is based upon work supported by the Department of Energy (DOE) under Grant No. DE-FG02-10ER41310, NASA under Grant No. NNX09AJ55G, and that part of the calculations for this work have been performed on the Cosmology Computer Cluster funded by the Hoblitzelle Foundation.  J.D. acknowledges that this research was supported in part by the DOE Office of Science Graduate Fellowship Program (SCGF). The DOE SCGF Program was made possible in part by the American Recovery and Reinvestment Act of 2009.  The DOE SCGF program is administered by the Oak Ridge Institute for Science and Education for the DOE. ORISE is managed by Oak Ridge Associated Universities (ORAU) under DOE Contract No. DE-AC05-06OR23100.}

\end{document}